\documentclass[aps,prb,floatfix,superscriptaddress,twocolumn,nofitinbib,hyperref=pdftex,citeautoscript]{revtex4}
\usepackage{epsfig}
\usepackage{amsmath,bm}
\usepackage{amssymb}
\usepackage{braket}
\usepackage{amsfonts}
\usepackage{dsfont}
\usepackage{comment,color}
\usepackage{lipsum}
\usepackage{hyperref}

\hypersetup{colorlinks=true,linkcolor=blue,anchorcolor=blue,citecolor=blue,filecolor=blue,urlcolor=blue,bookmarksnumbered=true,pdfview=FitB}

\newcommand{\nn}{\nonumber\\}
\usepackage[dvipsnames]{xcolor}

\begin{document}
\title{Tunable Josephson diode effect in singlet superconductor-altermagnet-triplet superconductor junctions}

\author{Lovy Sharma}
\affiliation{Department of Physics, Indian Institute of Technology Delhi, Hauz Khas, New Delhi, India 110016}
\author{Manisha Thakurathi}
\affiliation{Department of Physics, Indian Institute of Technology Delhi, Hauz Khas, New Delhi, India 110016}
\affiliation{Department of Physics, Indian Institute of Technology Hyderabad, Kandi, Sangareddy, Telengana, India 502285}
\date{\today}
\begin{abstract}
Recently discovered phase of collinear magnet called, altermagnet breaks time reversal symmetry (TRS), exhibits momentum-dependent spin-splitting of band structure with zero net magnetization. In this work, we theoretically investigate the Josephson junction (JJ) of spin singlet superconductor (SC)/altermagnet/spin triplet SC and demonstrate that it manifests field free Josephson diode effect (JDE). We illustrate that there are four key requisites to have JDE in such JJs, namely, broken TRS, left and right SC in the JJ shall be non-identical, presence of spin orbit interaction, and anisotropy in spin polarization at the Fermi surface or anisotropy in pair potential of the SC. It has also been shown that by applying a gate potential in the altermagnetic regime, one can not only reverse the sign of efficiency but also modulate its magnitude. Our system can be used as a superconducting rectifier that can be tuned efficiently using gate voltage and system parameters without having external magnetic field.
\end{abstract}
\maketitle

\section{Introduction}

The $p-n$ junction diode has revolutionized electronics due to its application in LEDs, photodiodes, voltage regulators, switching devices, rectifiers and many more [\onlinecite{6773080, 6372252}]. Recently, the superconducting diode effect (SDE) has drawn a lot of attention, where the critical current in one direction is different from the other [\onlinecite{Ando2020, PhysRevB.49.9244, Miyasaka_2021, PhysRevB.107.224518,Zhang2020-af,Pal2022-kq,PhysRevB.108.214520,PhysRevB.108.174516}]. This non-reciprocity leads to non-dissipative current in one direction while dissipative in the opposite, thus giving rise to energy-efficient counterpart of diode for quantum circuits. Theoretical studies and experimental realizations have demonstrated  SDE in bulk superconductors and Josephson diode effect (JDE) in Josephson junctions (JJs) [\onlinecite{ Narita2022-oi,Baumgartner2022-gt, PhysRevLett.131.196301,PhysRevX.12.041013, PhysRevB.106.214524,PhysRevLett.131.096001,PhysRevLett.128.037001,PhysRevLett.129.267702,Wu2022-cp,AS}]. Notably, breaking of inversion and time reversal symmetry (TRS) is generally required for the diode effect. In general, TRS is broken either by an external magnetic field or ferromagnet in the system. However, the presence of external magnetic field or ferromagnet has detrimental effect on the superconducting state which hinders its application to logical devices. Therefore, systems with field free JDE are of great interest and there have been some  theoretical and experimental studies where external magnetic field or ferromagnet is not required to break TRS [\onlinecite{Wu2022-cp,Lin2022-ok,PhysRevB.110.014518,debnath2024}]. JDE has been observed in van der Walls hetrostructure of $NbSe_{2}/Nb_{3}Br_{8}/NbSe_{2}$ junction without any magnetic field [\onlinecite{Wu2022-cp}], in JJ with chiral quantum dots [\onlinecite{debnath2024}], and it can occurs from finite Cooper pair momentum in a type-II topological semimetal  [\onlinecite{Pal2022-kq}]. Moreover, SDE is present in mirror symmetric twisted trilayer graphene without any external magnetic field due to the imbalance in the valley occupation of the Fermi surface [\onlinecite{Lin2022-ok}]. 
\begin{figure}[h!]
\centering
   \includegraphics[width=0.9\linewidth]{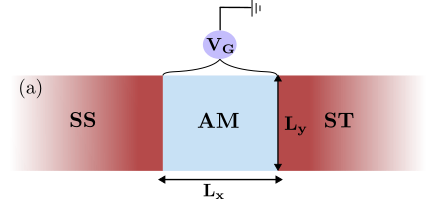}
     \vspace{1pt}
    \includegraphics[width=0.9\linewidth]{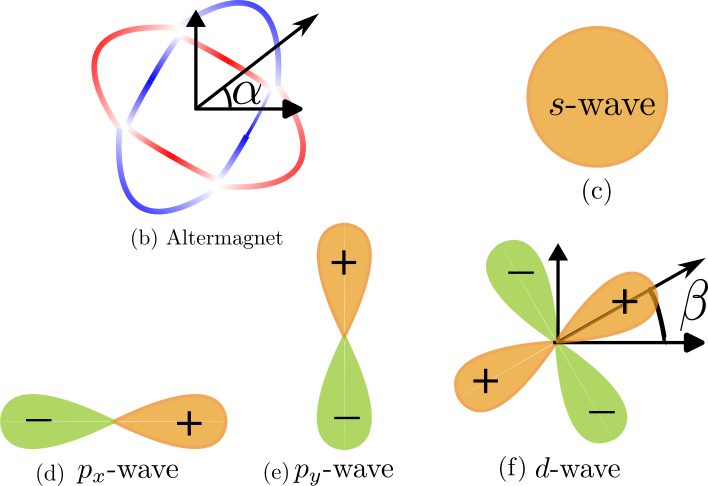}
    \caption{(a) Schematic diagram of SS/AM/ST Josephson junction having semi-infinite SCs in $x<0$ ($x>L_x$) with SS (ST) pairing symmetry. The junction is finite along $y-$axis with length $L_y$ and $N_y=L_y/a$ number of lattice points. The length of AM region is $L_x$ which has been biased using a gate potential $V_G$. (b) The Fermi surface of the AM with Rashba SOC, color grading represents spin orientation along (opposite) $z$-axis with red (blue), illustrating that the spin orientation changes sign around Fermi surface. (c)-(f) show the pairing function of $s$-wave, $p_x$-wave, $p_y$-wave and $d$-wave SCs respectively with positive and negative sign showing Cooper-pair phase change.}
\label{fig1} 
\end{figure}

Recently, a new class of magnetic material known as altermagnet (AM), with a collinear compensated magnetization, has been discovered [\onlinecite{PhysRevX.12.040002, PhysRevX.12.040501, PhysRevX.12.031042, PhysRevX.12.011028}]. AMs have net zero magnetization with $d-$wave magnetic order and  alternating spin polarization in both real as well as momentum space. AMs break TRS and exhibit spin-polarized Fermi surface like ferromagnets however have zero magnetization like antiferromagnets. The opposite-spin sublattice is connected by the same rotation symmetry in real and momentum space, wherein antiferromagnets it is connected by translation and inversion symmetry in real space, which is defining feature that differentiates them from both types of magnetism. The coexistence of AM and SC has been studied  using mean field theory, showing it favors mixture of $s-$wave and $p-$wave pairing symmetry [\onlinecite{PhysRevB.108.184505}]. The hetrostructure of SC and AMs have been found to accommodate first and second order topological superconductivity[\onlinecite{PhysRevLett.133.106601}]. A net zero magnetization renders AMs ideal candidates to break TRS without magnetic field.  Moreover, using AM to break TRS one can bypass the problem of stray fields in its application to quantum circuits. In our work, we utilize this property of AM to obtain JDE and illustrate that in junctions spin-singlet (SS) SC/AM/spin triplet (ST) SC, breaking TRS and inversion are not the sufficient conditions to observe JDE as some symmetries of the system prohibit it to have non-reciprocal effect. We emphasize that a $0-\pi$ transition has been studied in a JJ with SS SC/AM/ST SC [\onlinecite{PhysRevB.109.024517}], where TRS and inversion both are broken but the current phase relation (CPR) is still symmetric due to the symmetry satisfied by the system. In this work, we show that Rashba spin orbit coupling (SOC)[\onlinecite{PhysRevB.108.184505}] breaks those symmetries and results in a JDE where the efficiency reaches about $44\%$.  We use  $s$-wave and $d$-wave spin SS SC and $p-$wave ST SC to form the JJ. We show that for a wide range of system parameters,  efficiency of JDE of more than $20\%$. This efficiency changes sign, is tunable with the gate potential and stable against disorder.  We have also done a comparative study of junction SS SC/ferromagnet (FM)/spin ST SC. Notably, presence of SOI does not necessarily gives rise to JDE. We point out four requisites to obtain JDE in such JJs- first: broken TRS,  second: left and right SC in the JJ shall be``non-identical" (``non-identical" also includes SC with different angle of lobe direction of pair potential such as in d-wave SC), third: presence of SOI, and fourth: anisotropy in spin polarization at the Fermi surface or anisotropy in pair potential of SC. \\
The paper is organized as follows, in Sec. \ref{sec2} we present the model of our system and recursive Green's function method for computing current. Then, numerical results are shown in Sec. \ref{sec3} for the junction made up to $s$-wave and $d$-wave SC as SS superconductor. Variation of efficiency with system parameters on the basis of symmetry argument is explained in Sec. \ref{sec4}. In Sec. \ref{sec5} effect of the gate potential on efficiency has been elaborated, followed by a discussion of more similar systems in Sec. \ref{sec6}. And finally in Sec. \ref{sec7}, we conclude our results.

\section{Model and Formalism}
\label{sec2}
\subsection{Model}
We consider a two-dimensional planer Josephson junction in the $xy$-plane as shown in Fig.\ref{fig1} with SS SC/ AM/ ST SC. The junction is finite along the $y$ direction with length $L_y$ and the interface is formed along the $x$-axis with length $L_x$. Furthermore, the SC at the left end ($x<0$) and at the right end ($x>L_x$) are semi-finite. The system with operator $C^{\dagger}(k)= (\hat{c}^{\dagger}_{k,\uparrow} , \hat{c}^{\dagger}_{k,\downarrow}, \hat{c}_{-k,\uparrow}, \hat{c}_{-k,\downarrow})^{T}
$, is described by  Bogoliubov–de Gennes (BdG) Hamiltonian Hamiltonian $H_{SS/ST/AM}=C^{\dagger}(k) \hat{H}_{SS/ST/AM}(k) C(k)$ with $\hat{H}_{SS/ST}=\hat{H}_{0}(k) + \hat{\Delta}_{SS/ST}(k),
$ and, $\hat{H}_{AM}= \hat{H}_{0}(k) + \hat{J} (k)+\hat{\Lambda}(k)$. The kinetic energy ($\hat{H}_{0}(k)$), superconducting energy gap for SS ($\hat{\Delta}_{SS}(k)$) and ST ($\hat{\Delta}_{ST}(k)$), spin-orbit coupling ($\hat{\Lambda}(k)$), and d-wave nature AM $\hat{J} (k)$ terms take following form, respectively,
 
 \begin{align}
   & \hat{H}_{0}= [t_{0}(k_{x}^{2}+k_{y}^{2}) - (\mu+V_G)]\tau_{z}\otimes\sigma_{0}, \label{eq1}\\
&    \hat{\Delta}_{SS}=
\begin{pmatrix}
    0 & \Delta(k) e^{-i\phi_{L}}i\sigma_{y}\\
     -\Delta^{*}(-k) e^{i\phi_{L}}i\sigma_{y} & 0
\end{pmatrix},\\
&\hat{\Delta}_{ST} = \begin{pmatrix}
  0 & d(k)\sigma_{x}\\
  -d^{*}(-k)\sigma_{x} & 0
\end{pmatrix},\\
&\hat{\Lambda}=\lambda[(k_{y}\cos{\alpha}-k_{x}\sin{\alpha}) \tau_{0}\otimes\sigma_{x}-\nonumber\\
&(k_{x}\cos{\alpha}+k_{y}\sin{\alpha}) \tau_{z}\otimes\sigma_{y}], \label{eq4}\\
&   \hat{J}=J_{a}\big[ (k_{x}^{2}-k_{y}^{2})\cos{2\alpha} + 2k_{x}k_{y}\sin{2\alpha}\big]\tau_{z}\otimes\sigma_{z}.
 \label{eq2}
\end{align}
The Pauli matrix $\tau$ ($\sigma$) acts in the particle-hole (spin) space. Here, the parameters $t_{0}$, $\mu$, $J_a$, $\lambda$, $\Delta_{0}$ and $\alpha$ are the strengths of hopping, chemical potential, AM, Rashba SOC, SC pairing amplitude and crystallographic angle of AM with the $x$-axis ( as shown in Fig.\ref{fig1}), respectively. The phases $\phi_{L}$ and $\phi_{R}$ are macroscopic phases of the left and right SC. Moreover, the phase difference between the left and right SCs is taken as $\phi=\phi_{R}-\phi_{L}$. The form of gate potential is $V_G=V [\Theta(x)-\Theta(x-L)]$ as shown in Fig. \ref{fig1}.  The gap parameter $\Delta(k)=\Delta_{0}$ for $s$-wave SC and $\Delta(k)=\Delta_{0}((k_{x}^{2}-k_{y}^{2})\cos{2\beta} + 2k_xk_y\sin{2\beta}$ for $d$-wave SC where $\beta$ denotes the angles at which the lobe direction of the pair potential orients with respect to the $x$-axis as shown in Fig.\ref{fig1}. In ST SC, the form of $d(k)=\Delta_{0}(\eta_{1}k_{x}+ i \eta_{2}k_{y}) \exp(-i\phi_{R})$ with $(\eta_{1},\eta_{2})=(1,0)$, $(0,-i)$, and $(1,1)$ for $p_x$, $p_y$ and chiral p-wave pairing of SC, respectively.  

In the later section, we will analyze the system using recursive Green's function algorithm to find the current phase relationship (CPR), for which the real space tight-binding Hamiltonian of the system is required. Therefore, we model a junction of width $L_y$, such that the $N_y=L_y/a$ number of lattice points along the $y$ direction, $0<x<L_x$ is the region with AM which has been biased using gate potential $V_G$ and $x\leq0$ ($x\geq L_x$) is for left (right) SC.  The discretized  Hamiltonian for the system has following form

\begin{align}
H=\hat{H}^{LS} + \hat{H}^{AM} + \hat{H}^{RS}+
\hat{H}^{C},
\label{Eq6}
\end{align}
where, $\hat{H}^{LS}$, $\hat{H}^{AM}$, $\hat{H}^{RS}$ are tight binding Hamiltonian in left SC, AM and right SC,  respectively. Moreover,  $\hat{H}^{C}$ contains two hopping parts: first between the left SC and the AM ($\hat{H}_{tL}$) and the second one between the AM and right SC ($\hat{H}_{tR}$). The forms of these matrices are written in Appendix \ref{AppA}.

\subsection{Current Formalism}
In this section, we briefly review the Green's function method to compute the Josephson current across the junction. The current phase relation (CPR) can be calculated using the number operator in the left SC [\onlinecite{PhysRevB.110.014518,PhysRevB.104.134514,Sun_2009,PhysRevB.93.195302,doi:10.7566/JPSJ.83.074706,PhysRevResearch.2.023197}], $\hat{N}=\sum_{x\leq0,y,s} \hat{\psi}^{\dagger}_{x,y,s}\hat{\psi}_{x,y,s}$ where $s=(\uparrow,\downarrow)$. as,
\begin{align}
    &\big<I\big>=e\Big< \frac{d\hat{N}}{dt}\Big>=\frac{ie}{\hbar}\Big<[H,\hat{N}] \Big>,\\
    &\big<I\big>=-\frac{e}{h}\int dE Tr[\Gamma_{z}\hat{H}_{tL}G^{<}_{LJ}(E)+\text{H.c.}].
    \label{13}
\end{align}
Here, we have divided the system into strips of length $N_y a$ which are placed along $x-$direction.  The matrix $\Gamma_{z}=\sigma_{z}\otimes I_{2}$ and {$\hat{H}_{tL}$ (defined in Appendix \ref{AppA}) is the first part of hopping matrix $\hat{H}^C$ written in Eq. (\ref {Eq6})  which connects left SC at $x=0$ and AM at $x=1$ strip. The non-local lesser Green's function $G^{<}_{LJ}(E)$ is defined at the left junction (LJ) i.e. between $x=0$ and $x=1$ strip and is computed via Fluctuation-Dissipation theorem as
\begin{align}
    G^{<}_{LJ}(E)=-f(E)[G^{r}_{LJ}(E)-G^{a}_{LJ}(E)],
\end{align}

\begin{figure}[b]
\centering
\begin{tabular}{c c}
   \includegraphics[width=0.5\linewidth]{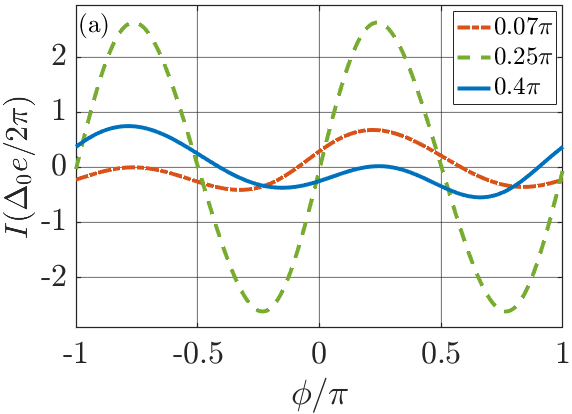}
& \hspace*{-2pt}\includegraphics[width=0.49\linewidth]{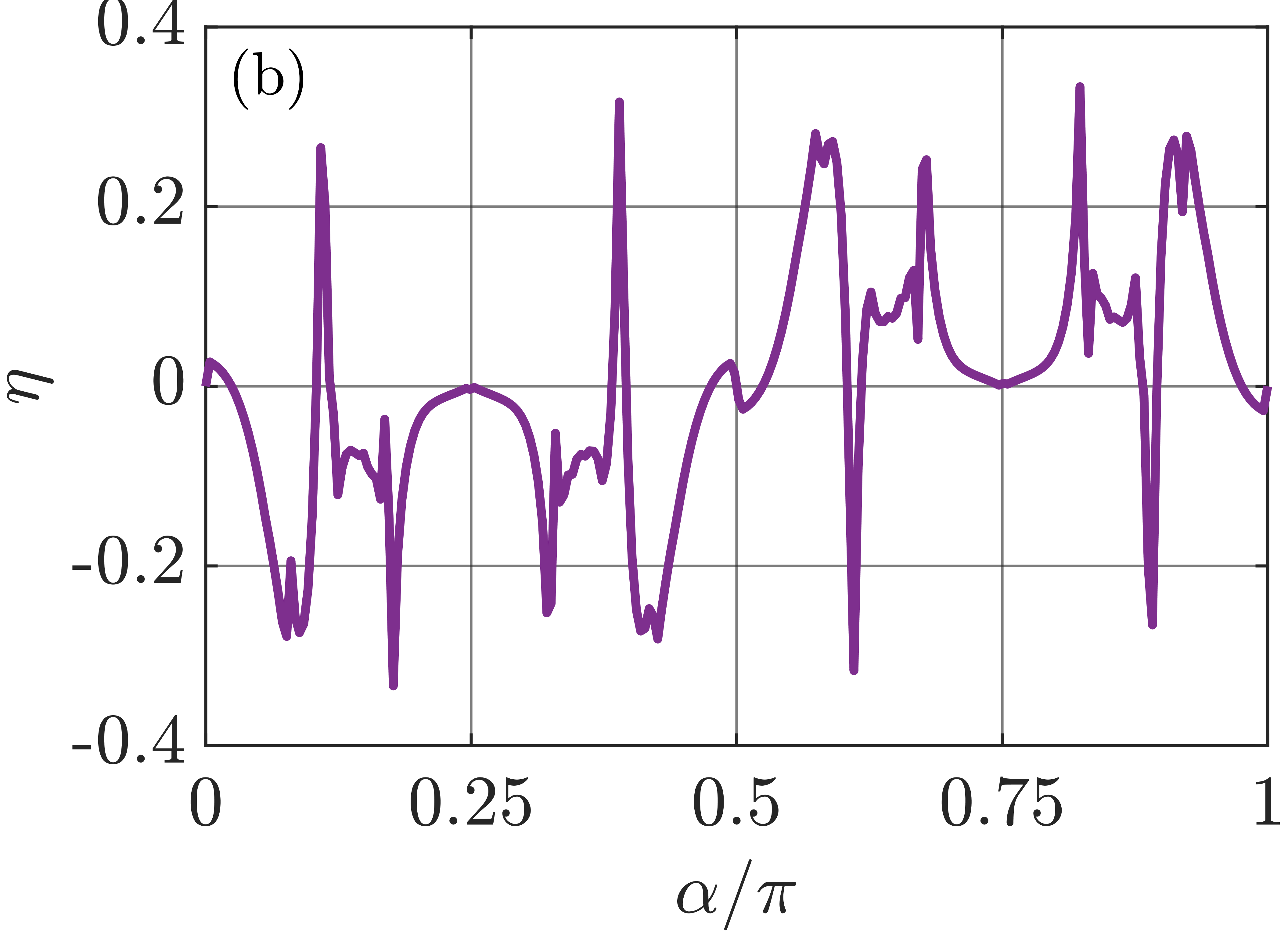}
    \end{tabular}
    \caption{Plot of CPR and $\eta$ variation for $s$-wave SC/AM/$p_x$-wave SC JJ (a) CPR for $\alpha=0.07\pi$, $0.25\pi$, and $0.4\pi$. (b) variation of $\eta$ as a function of $\alpha$ for $J_a=0.7$, $\lambda=0.2$, $N_x=4$ and $N_y=6$.}
\label{fig2} 
\end{figure}

where $f(E)$ is the Fermi distribution function. The retarded (advance) Green's function can be obtained using the relation $G^{r(a)}_{LJ}(E)=G^{r(a)}_{AM}(E)\hat{H}_{tL}^{\dagger}g^{r(a)}_{LS}(E)$, where $g^{r(a)}_{LS}$ is the surface Green's function of left semi-infinite SC and computed using M\"obius transformation, the methodology is explained in Appendix \ref{mob}. 
 Moreover, $G^{r(a)}_{AM}$ is the retarded (advanced) Green's function in AM at $x=1$ strip which can be calculated using recursive algorithm as follows. We first note that inside the AMic region, the on-strip Hamiltonian (denoted as $H_{11}^{AM}$) and hopping  Hamiltonian (denoted as $H_{12}^{AM}$) from one strip to the next right strip are independent of the strip label (junction boundaries are not involved). We begin the algorithm from the right end, the surface Green's function of the right SC ($g^{r(a)}_{RS}$) is obtained using M\"obius transformation. Now for the right most strip in AM at $x=N_x$ the retarded Green's function is written as,
\begin{align}
    G^{r(a)}(E,N_{x})=\big(E-H_{11}^{AM}- \hat{H}_{tR}^{\dagger} g^{r(a)}_{RS} \hat{H}_{tR}\big)^{-1}.
\end{align}
 
 Here,  $\hat{H}_{tR}$ is defined in Appendix \ref{AppA} which connects AM at $N_x$th strip to SC at $(N_x+1)$th strip. In the next block, the retarded Green's function at $x=(N_x-1)$ takes the form
\begin{align}
    G^{r(a)}&(E,N_{x}-1)=\nn
    &\big(E-H_{11}^{AM}- H_{12}^{AM} G^{r(a)}(E,N_{x}) H_{12}^{AM\dagger} \big)^{-1}.
\end{align}
 We continue aforementioned process upto $x=1$ and obtain $G^{r(a)}(E,1)$. However, $G^{r(a)}_{AM}$ has the contribution from the surface Green's function of left SC as well in the algorithm, therefore, it has the following form. 
\begin{align}
    G^{r(a)}_{AM}(E)=\big((G^{r(a)}(E,1))^{-1}  - \hat{H}_{tL}^{\dagger} g^{r(a)}_{LS} \hat{H}_{tL}\big)^{-1},
\end{align}
where the form of $(G^{r(a)}(E,1))^{-1}=E-H_{11}^{AM}- H_{12}^{AM} G^{r(a)}(E,2) H_{12}^{AM\dagger}$. Hence,  we have everything required to calculate CPR numerically.

\begin{figure*}
\centering
\begin{tabular}{c c c}
   \includegraphics[width=0.287\linewidth]{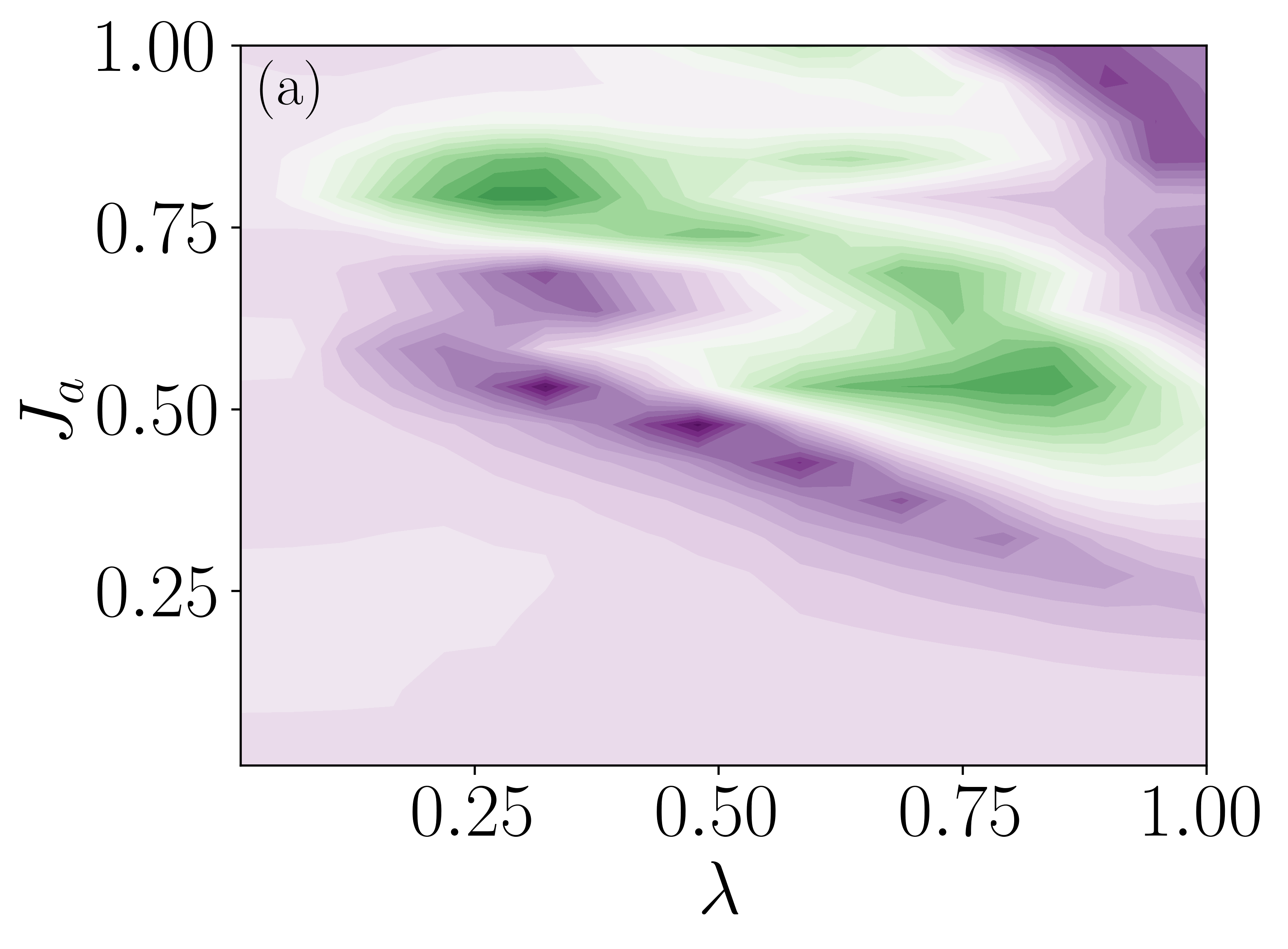}
& \includegraphics[width=0.27\linewidth]{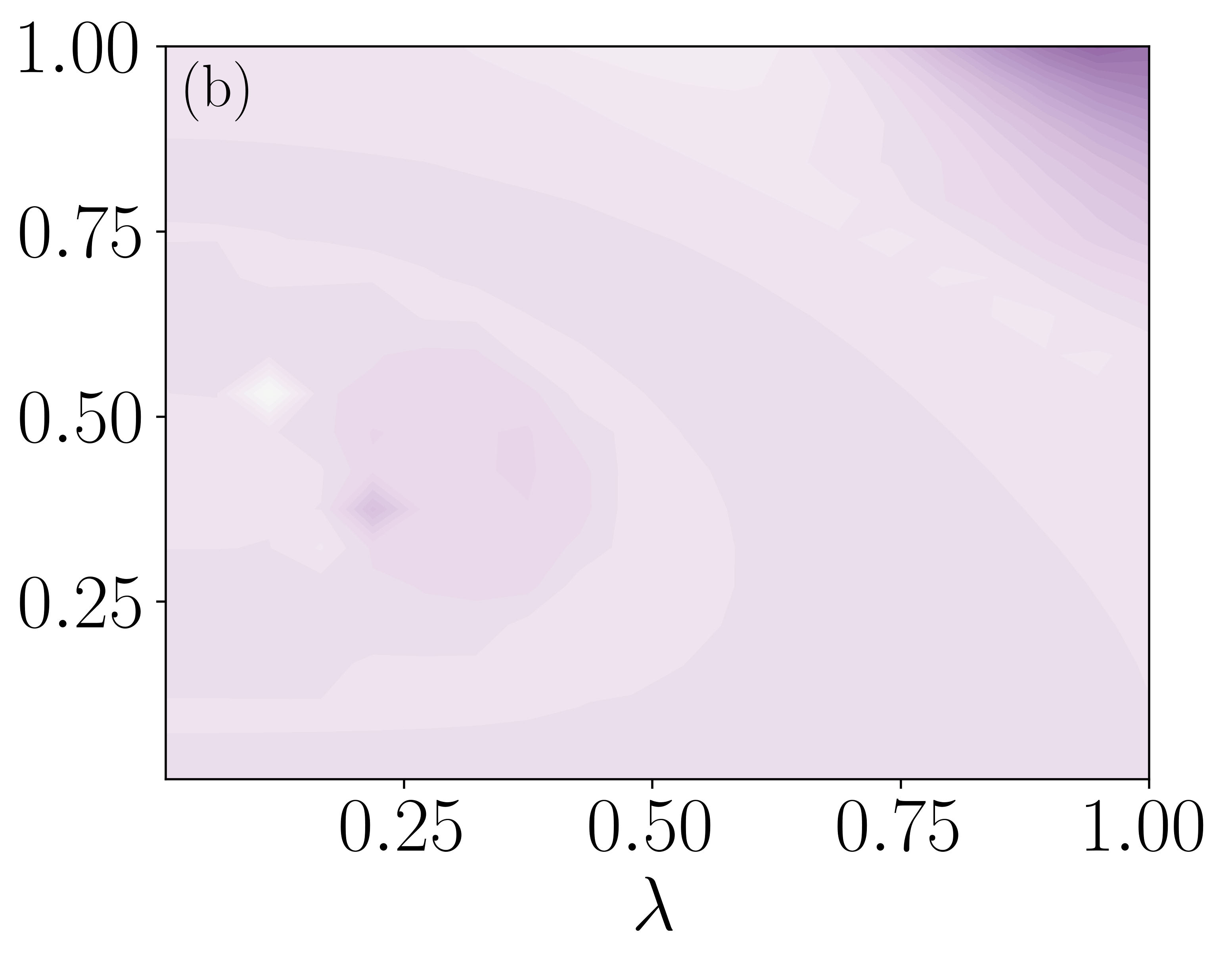}
&\includegraphics[width=0.33\linewidth]{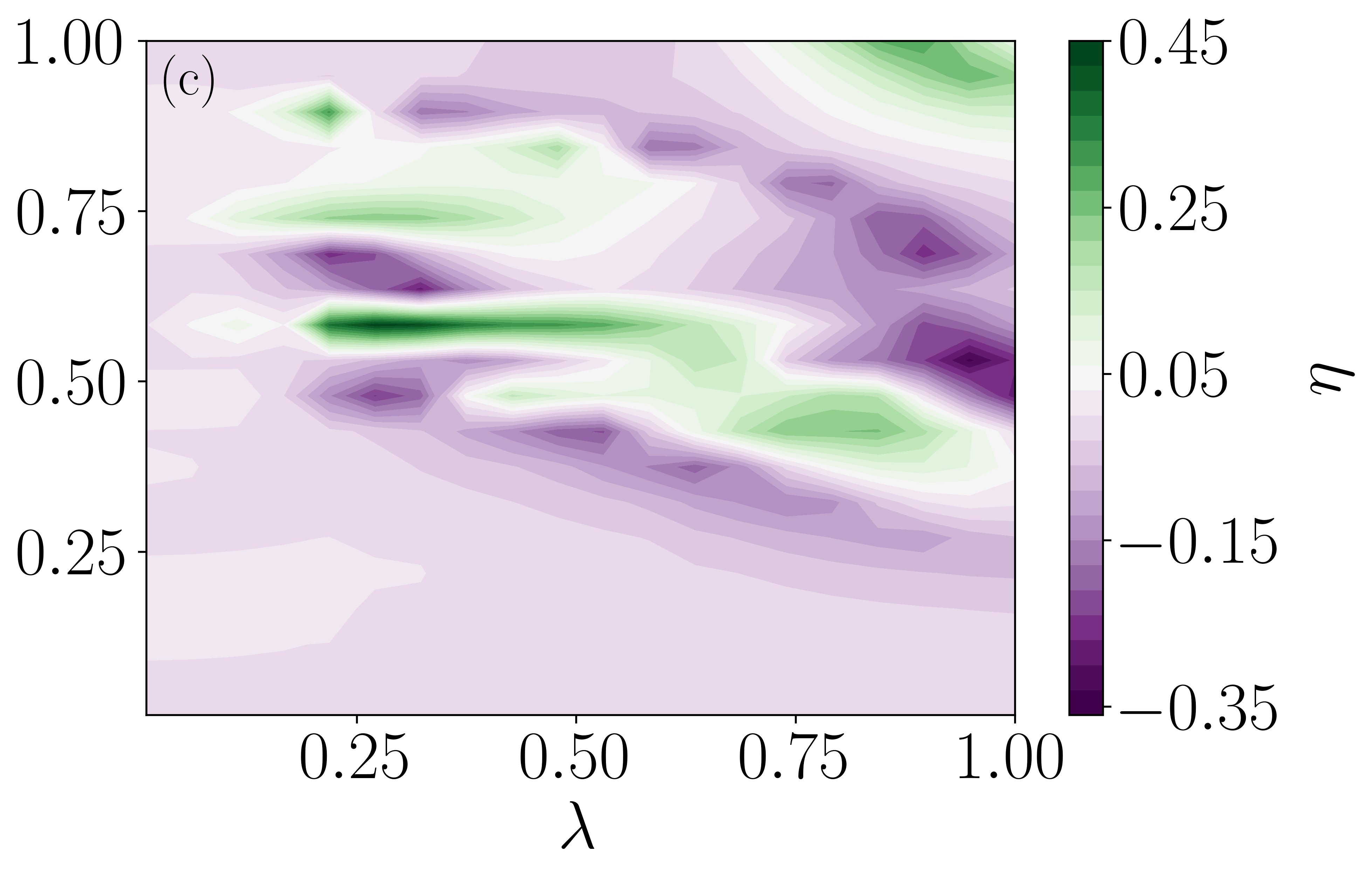}
    \end{tabular}
    \caption{(a-c) Contour plot of $\eta$ w.r.t $J_a$ and $\lambda$ for (a)$\alpha=3\pi/8$, (b)$\pi/4$ and (c)$0.07\pi$ respectively. with $N_x=4$ and $N_y=6$  }
\label{fig3} 
\end{figure*}

\section{Numerical Results}
\label{sec3}
In this section, we numerically compute CPR using Eq. (\ref{13}). We take $0<J_a<\mu/2$ in our analysis for the well-defined Fermi surface [\onlinecite{PhysRevLett.133.226002}], $0<\lambda<\mu/2$,  $\mu=2$, hopping strengths $t_0=t=1$ and $\Delta_{0}=0.01$ throughout the paper, unless mentioned specifically. The number of lattice points, $N_x=4$ and $N_y=6$ are such that we are in the short junction regime since the coherence length of our system ($\hbar v_{f}/\Delta_{0}$) is much larger than $N_x$.
First, we analyze $s$-wave SC/ AM/ $p_x$ SC junction. The CPR for different values of $\alpha=0.07\pi$, $\pi/4$ and $0.4\pi$ is shown in Fig.\ref{fig2}a. The current is normalized with $e\Delta_{0}/2\pi$. For each CPR plot, the extrema of the supercurrent in both positive  ($I_c^{+}$) and negative directions ($I_c^{-}$) have been computed such that $I_c^{+}=\text{max}(I(\phi))$ and $I_c^{-}=\text{max}(-I(\phi))$ respectively with $0<\phi<2\pi$.  The CPR plots for $\alpha=0.07 \pi$ and $0.4 \pi$ clearly illustrate that the $I_c^{+}$ and $I_c^{-}$ are of different magnitudes. To quantify non-reciprocity, we define  the Josephson diode efficiency (JDE), $\eta=(I_c^+-I_c^-)/(I_c^++I_c^-)$ which takes values $0.24$ ($0.15$) for $\alpha=0.07 \pi$ ($\alpha=0.4 \pi$). We note that $\alpha=0.25 \pi$  is an special case where CPR is symmetric with $\sin(2\phi)$ type phase relation [\onlinecite{PhysRevB.109.024517}] in contrast with other values of $\alpha$  where $\sin(\phi)$, $\cos(\phi)$ and $\sin(2\phi)$ dependency coexist in CPR to facilitate the non-reciprocity. This can also be understand using the symmetry arguments given in Sec. \ref{sec4} where the functional form of the current, due to the symmetry transformation, prohibits one or other kind of sinusoidal functions, and hence non-reciprocity. It is evident from Fig. \ref{fig2}(b) that the JDE oscillates between positive and negative values as the crystallographic angle of AM, i.e. $\alpha$ varies from zero to $\pi$.   To further analyze  the dependence of $\eta$ on $J_a$ and $\lambda$, we have plotted contours of $\eta$ for different lobe angles of AM  as shown in Fig. \ref{fig3}.   In Fig. \ref{fig3}(a), for parameters $J_a=0.9$ and $\lambda=0.2$ the value of eta is about $30\%$ where $\lambda<J_a$. Also, for $J_a\approx0.6$, $\eta>28\%$ exist for $0.2<\lambda<0.5$, similar range of variation for $0.45<J_a<0.7$ for $\lambda\approx0.3$, (which is within the experimentally founded values of $J_a$ [\onlinecite{PhysRevX.12.011028}]). The maximum efficiency is approx $31\%$ around $J_a=0.8$ and $\lambda\approx0.27$. However, the efficiency for more than $20\%$ exists for very large region of $J_a$ and $\lambda$. For the chiral-$p$ wave SC on the right, the variation of $\eta$ is similar to the one with $p_x$-wave SC, however the efficiency is slightly larger compared to the $p_x$-wave SC. A similar pattern is achieved by varying values $\alpha$  and hence we are showing the variation only for $p_x$-wave SC. Moreover, for the $p_y$-wave SC on the right, the efficiency is negligible thus reflecting the fact that only the $p_x$ component in SC is contributing to the non-reciprocity and thereby to the diode effect.

\begin{figure}[b]
\centering
\begin{tabular}{c c}
   \includegraphics[width=0.49\linewidth]{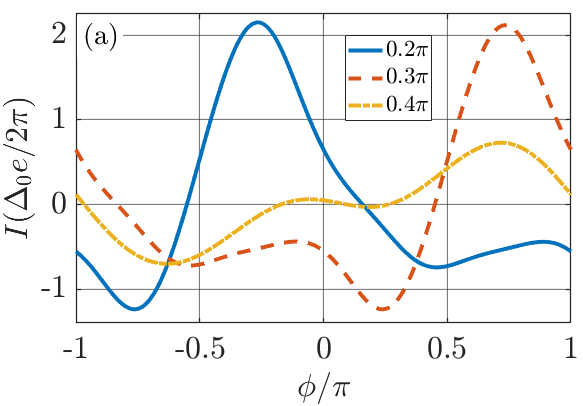}
& \includegraphics[width=0.5\linewidth]{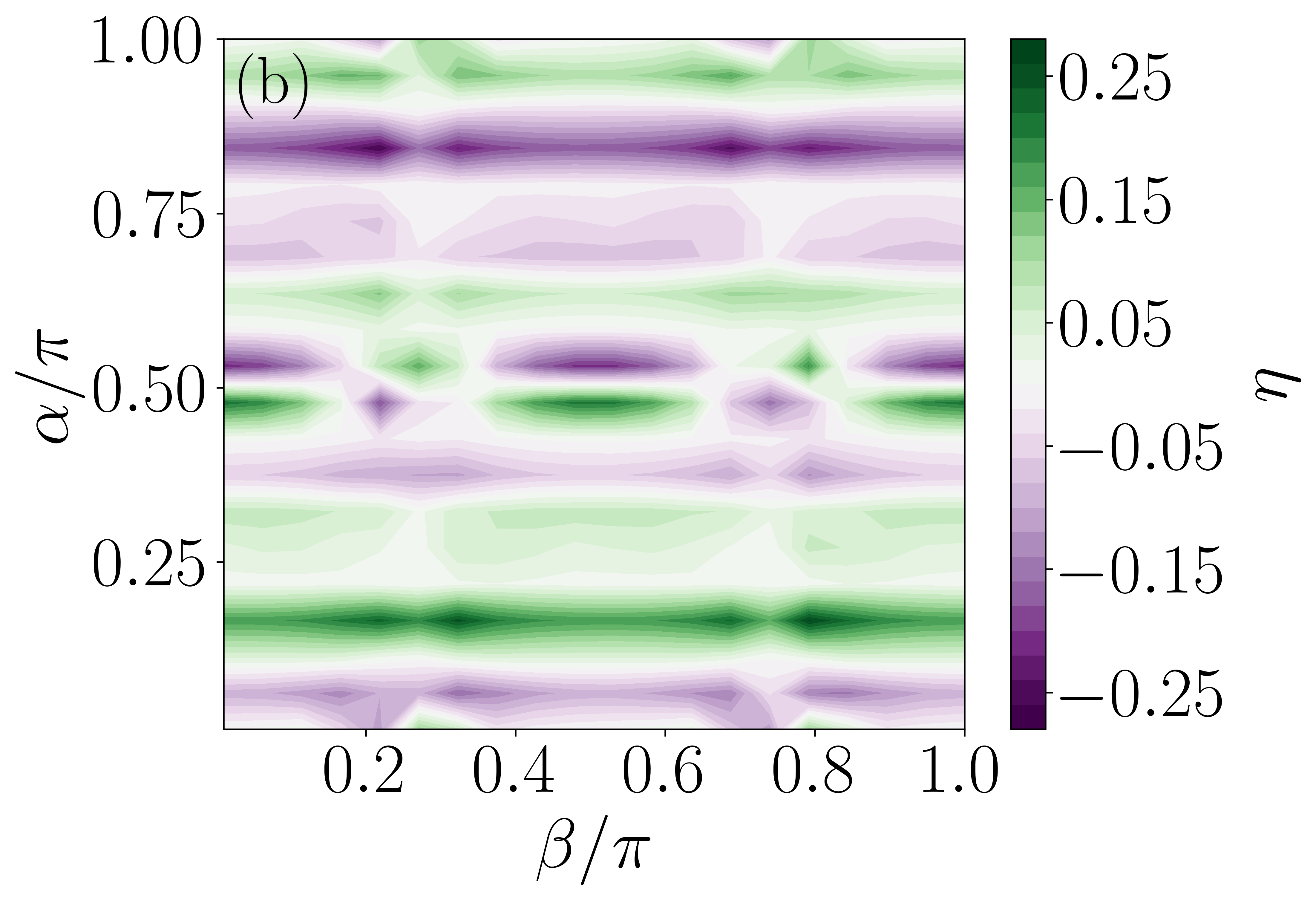}
    \end{tabular}
    \caption{CPR and $\eta$ of $d-$wave SC/AM/$p_x$-wave JJ for $N_x=4$ and $N_y=6$ (a) CPR with $J_a=0.7$, $\lambda=0.4$ and $\beta=0.45\pi$ for $\alpha=0.2\pi$, $0.3\pi$ and $0.4\pi$. (b) Countor plot shows the variation of $\eta$ as a function of crystallographic angle of AM $\alpha$ and lobe angle of pairing potential of $d$-wave SC, $\beta$ from $x-$axis for $J_a=0.6$ and $\lambda=0.3$. }
\label{fig4} 
\end{figure}

For $d_{xy}$ AM i.e., $\alpha=\pi/4$, efficiency above $10\%$ exists only for very high values of $J_a$ and $\lambda$, see Fig.\ref{fig3}(b). For $J_{a}$ and $\lambda$ less than 0.9, $\eta$ is almost insignificant which is also apparent in CPR for $J_a=0.7$ and $\lambda=0.2$ in Fig.\ref{fig2} and shows $\sin(2\phi)$ character similar to the case when $\lambda=0$ [\onlinecite{PhysRevB.109.024517}]. Moreover, for $\alpha=0$, diode effect is absent and slight deviation from this high symmetric point results $\eta$ to be non-zero for very rich ranges of parameters as shown for $\alpha=0.07\pi$ in Fig.\ref{fig3} (c).  Here, for parameter values $J_a\approx 0.6$, $0.2<\lambda<0.3$, the efficiency is $44\%$.

Next we analyze a $d-$wave spin singlet/AM/$p_{x}$-wave SC junction. The CPR is plotted for $J_a=0.7$, $\lambda=0.4$ and $\beta=0.45\pi$ for three different values of $\alpha$ in Fig.\ref{fig4} (a). Apart from $\alpha=0.4\pi$, the non-reciprocity is clearly evident and has a value of about $\eta=0.27$ for $\alpha=0.2\pi$ and $0.3\pi$. For the same parameters, diode efficiency is lower than that achieved with $s$-wave SC as SS. In Fig. \ref{fig4}(b), we have also plotted the variation of efficiency with  $\alpha$ and lobe angle of $d-$wave SC, $\beta$ with $J_a=0.6$ and $\lambda=0.3$. The maximum values of $\eta$ are found around $\alpha=0.15\pi$ and $0.85\pi$. Strips like pattern is obtained in this contour reflecting that efficiency does not change much on changing lobe angle of $d-$wave SC. These patterns in the graph can be understand using symmetry analysis presented in next section. The variation of $\eta$ with respect to $J_a$ and $\lambda$ is shown in the contour plots Fig.\ref{fig5}, where $\beta=0.45\pi$. For $\alpha=0.07\pi$ the maximum efficiency where $J_a>\lambda$ is about $28\%$ near $J_a=0.68$ and $\lambda=0.3$. Here the variation is smoother than that of $s-$wave SC counterpart but less richer in range of parameter. For $\alpha=\pi/4$ the efficiency is generally small and only has some notable values particularly for large values of $J_a$ and $\lambda$.

\begin{figure*}
\centering
\begin{tabular}{c c c}
   \includegraphics[width=0.287\linewidth]{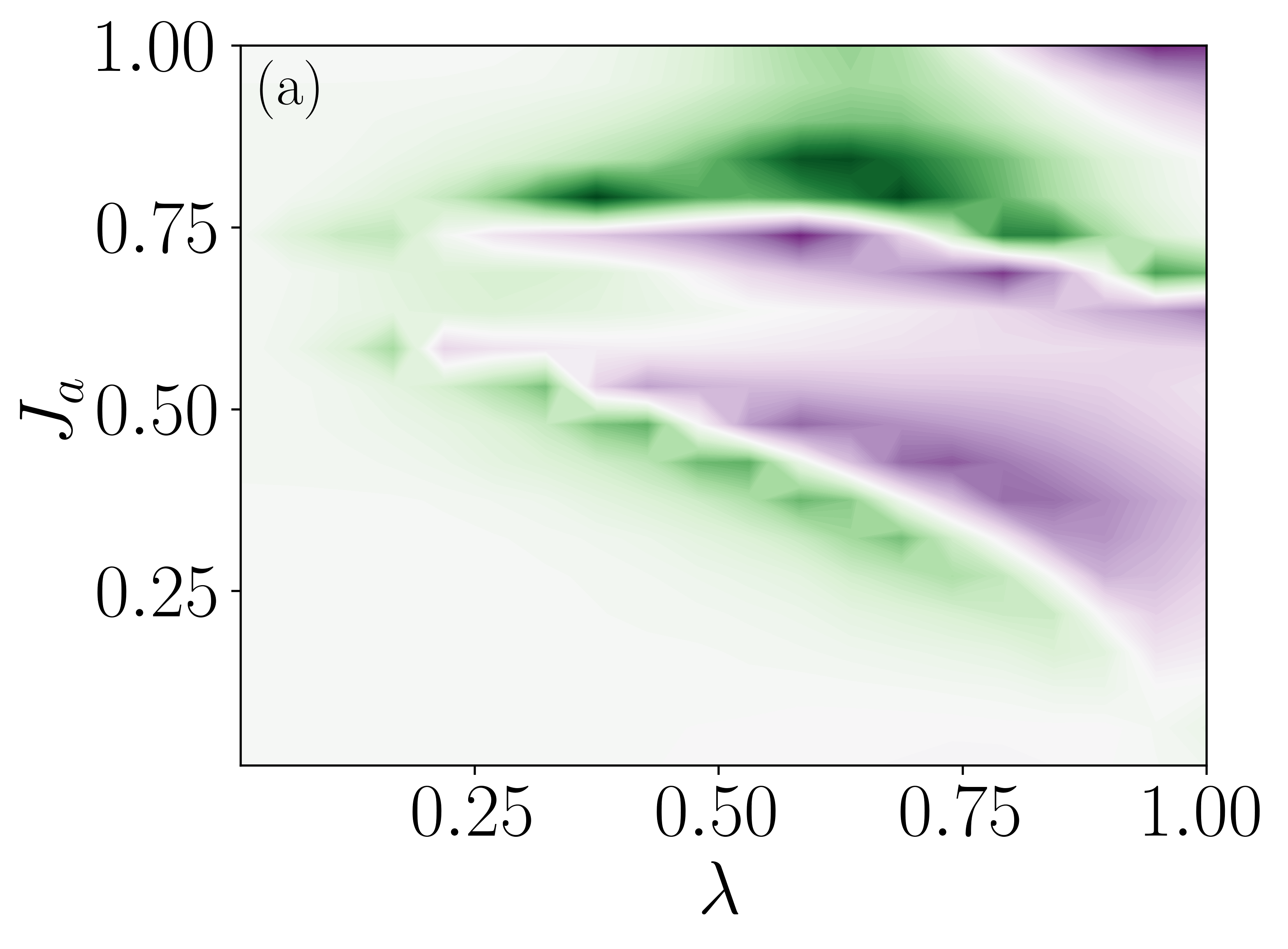}
& \includegraphics[width=0.27\linewidth]{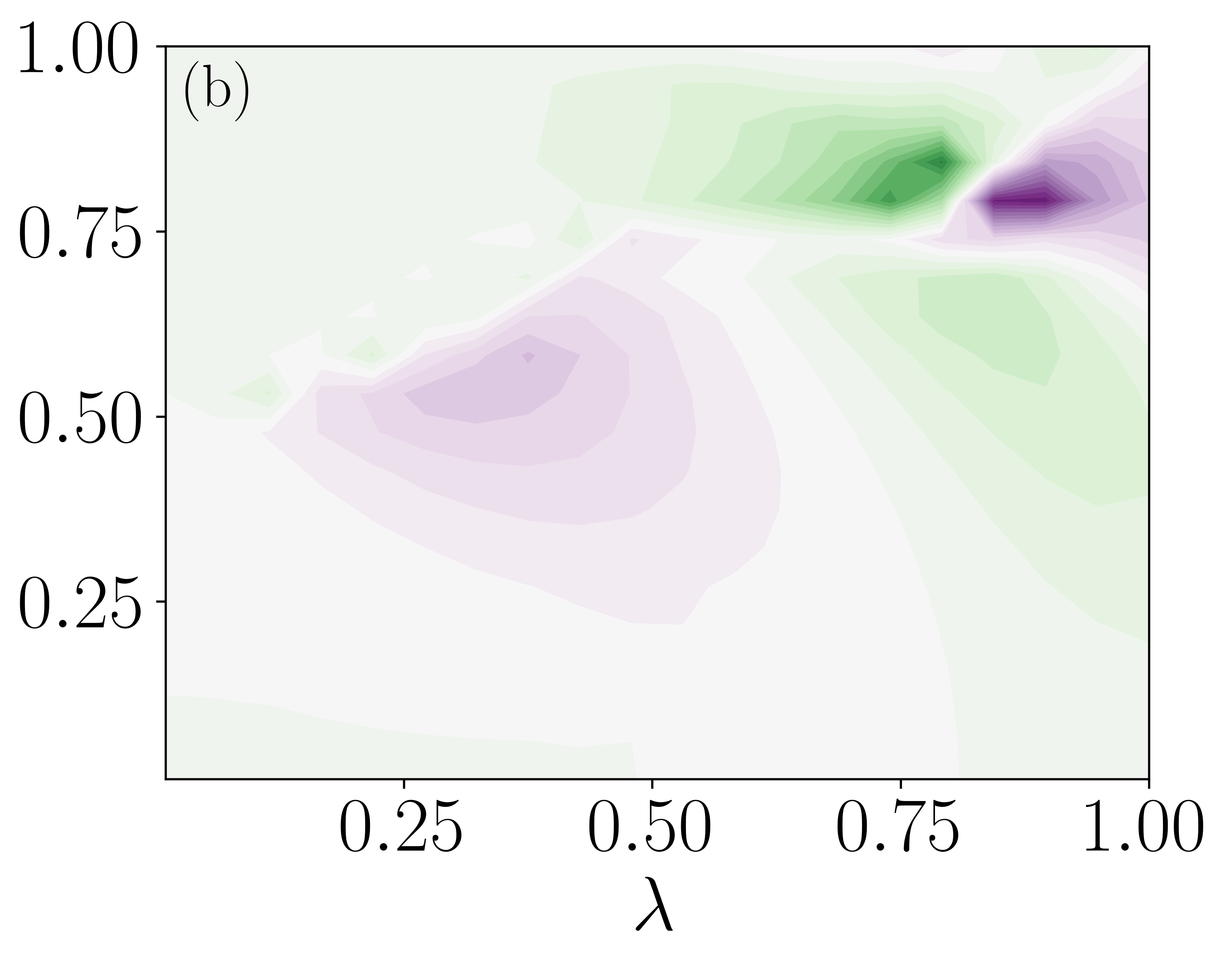}
&\includegraphics[width=0.32\linewidth]{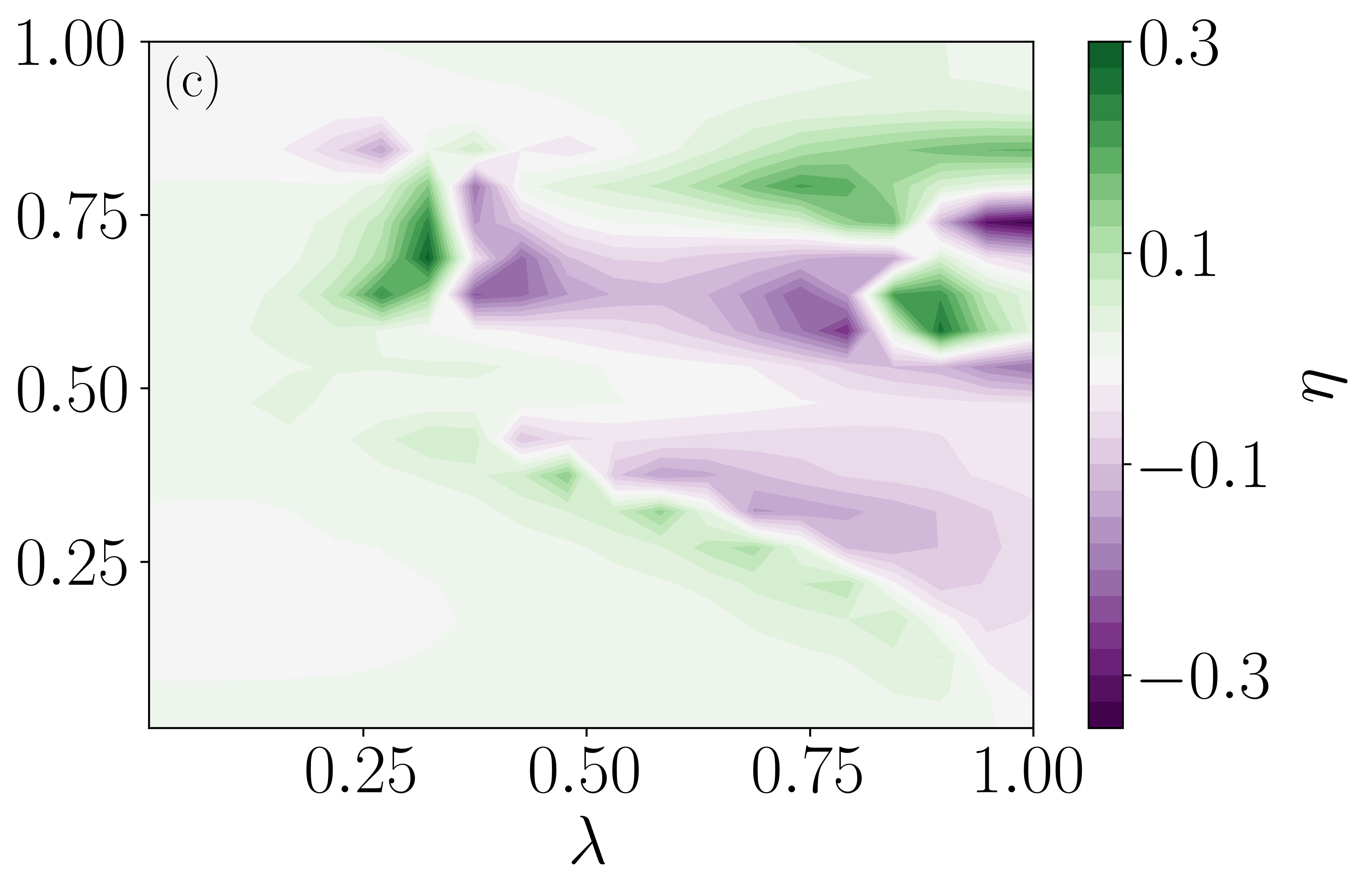}
    \end{tabular}
    \caption{(a-c) Contour plot of $\eta$ w.r.t $J_a$ and $\lambda$ where $\beta=0.45\pi$ for (a) $\alpha=3\pi/8$, (b) $\pi/4$ and (c) $0.07\pi$ respectively. }
\label{fig5} 
\end{figure*}

\section{Symmetry Annalysis}
\label{sec4}
In this section, we derive the symmetries of CPR using Hamiltonians written in Eqs. (\ref{eq1}-\ref{eq2}). First we show that even after breaking inversion and TRS using $p-$wave SC and AM, respectively, SOC is necessary for the non-reciprocal effect. Moreover, even with SOC the non-reciprocity is not present for some specific crystallographic angle of AM, which is evident in the plots of $\eta$ as a function of $\alpha$ shown in Fig. \ref{fig2} and Fig. \ref{fig4}. We first define four symmetry transformations: TRS ($\mathcal{T}$), mirror symmetry in $x-z$ plane ($\mathcal{M}_{xz}$), spin rotation of $\pi$ about $y-$axis ($\mathcal{R}_{y}$) and about $z-$axis ($\mathcal{R}_{z}$) [\onlinecite{PhysRevB.104.134514}]. The matrix form of these transformations is given by,
\begin{align}
    \mathcal{T}=\begin{pmatrix}
        -i\sigma_{y}&0\\
        0 & -i\sigma_{y}
    \end{pmatrix}\mathcal{K},
\end{align}

\begin{align}
    \mathcal{M}_{xz}=\begin{pmatrix}
        i\sigma_{y}&0\\
        0 & i\sigma_{y}
    \end{pmatrix}\mathcal{I}_{y},
\end{align}

\begin{align}
    \mathcal{R}_{y}=\begin{pmatrix}
        -i\sigma_{y}&0\\
        0 & -i\sigma_{y}
    \end{pmatrix},
\end{align}
\begin{align}
    \mathcal{R}_{z}=\begin{pmatrix}
        -i\sigma_{z} & 0\\
        0 & i\sigma_{z}
    \end{pmatrix},
\end{align}}
where $\mathcal{K}$ is complex conjugation transformation, $\mathcal{I}_{y}$ is the inversion of $y$-axis which results in $y\rightarrow-y$ hence $k_{y} \rightarrow -k_{y}$ and $\sigma_i$ denote the Pauli matrices in spin-space. We also define three joint operations using the aforementioned transformations: $\mathcal{X}=\mathcal{T}\mathcal{R}_{y}$, $\mathcal{Y}=\mathcal{R}_{z}\mathcal{T}\mathcal{M}_{xz}$, and $\mathcal{Z}=\mathcal{T}\mathcal{M}_{xz}$. Further we analyze the effect of $\mathcal{X}$ symmetry transformation on the Hamiltonian for $s$-wave SC/AM/$p_x$-wave SC junction without rashba SOC.
\[
\mathcal{X}\mathcal{H}_0(\phi_{L},\phi_{R},\alpha)\mathcal{X}^{-1}=\mathcal{H}_0(-\phi_{L},\pi-\phi_{R},\alpha),
\]
here, $\mathcal{H}_0= H_{SS}+H_{AM}(\lambda=0)+H_{TS}$. As $\mathcal{X}$ is combination of TRS and spin rotation, it will reverse the current direction. Therefore, with $\phi=\phi_R-\phi_L$, $I(\phi,\alpha)= -I(\pi-\phi,\alpha)$ this implies that $I_c^{+}$ will always be equal to $I_c^{-}$ hence no non-reciprocity. However, for a nonzero value of Rashba SOC in our system, $\mathcal{X}$   symmetry transformation is broken which allows non-reciprocal CPR in the system. Same transformation also prohibits non-reciprocity in the junction $d$-wave SC/AM/$p_x$-wave SC. Consequently, SOC is the necessary condition for the diode effect in our system.\\
Next, we explore the symmetry constraints along certain values of $\alpha$ and $\beta$ and the reasons for specific patterns in the graphs and contours. Consider the symmetry transformation $\mathcal{Y}$ that transforms the Hamiltonian, $\mathcal{H}= H_{SS}+H_{AM}+H_{TS}$ of  $s-$wave SC/AM/$p_{x}$-wave SC as
\begin{align}
    \mathcal{Y}\mathcal{H}(\phi_{L},\phi_{R},\alpha)\mathcal{Y}^{-1}=\mathcal{H}(-\phi_{L},\pi-\phi_{R},\pi-\alpha).
    \label{17}
\end{align}
This leads us to the current relation, $I(\phi,\alpha)=-I(\pi-\phi,\pi-\alpha)$ which explains the pattern of $\eta$ in Fig.\ref{fig2} (b) as follows. We note that currents $I_c^{+}(\alpha)=I_c^{-}(\pi-\alpha)$ therefore, as illustrated, $\eta$ for any lobe angle $\alpha$ has the same magnitude for lobe angle $(\pi-\alpha)$ with opposite sign. It also explains the reason for zero value of $\eta$ at $\alpha=\pi/2$. Further, at $\alpha=0$ and $\alpha=\pi$,  we obtain $\eta=0$ which can be explained using symmetry operation $\mathcal{Z}$ that transforms Hamiltonian of the $s-$wave SC/AM/$p_{x}$-wave SC junction as,
\begin{align}
     \mathcal{Z}\mathcal{H}(\phi_{L},\phi_{R}, \alpha=0,\pi)\mathcal{Z}^{-1}=\mathcal{H}(-\phi_{L},\pi-\phi_{R},\alpha=0,\pi).
\end{align}
This indicates $I(\phi)=-I(\pi-\phi)$ and hence for $\alpha=0$ and $\alpha=\pi$ the efficiency would be zero even though the SOC is present.\\

In the $d$-wave SC/ AM/ $p_x$-wave SC junction, $\mathcal{Y}$ transformation transforms Hamiltonian as,
\begin{align}
    \mathcal{Y}\mathcal{H}(\phi_{L},\phi_{R},\alpha,\beta)\mathcal{Y}^{-1}=\mathcal{H}(-\phi_{L},\pi-\phi_{R},\pi-\alpha,\pi-\beta).
    \label{eq19}
\end{align}
Again following the similar discussion on current, one can argue that the $\eta=0$ for $\alpha=\pi/2$ and $\beta=0,$ $\pi/4$, $\pi/2$ and $3\pi/4$ as follows. 
\begin{align}
\label{23}
    \mathcal{Y}\mathcal{H}(\phi_{L},\phi_{R},\alpha=\pi/2, \beta=0,\pi/2)\mathcal{Y}^{-1}=\nn \mathcal{H}(-\phi_{L},\pi-\phi_{R},\alpha=\pi/2, \beta=0,\pi/2),
\end{align} 
\begin{align}
\label{24}
    \mathcal{Y}\mathcal{H}(\phi_{L},\phi_{R},\alpha=\pi/2, \beta=\pi/4,3\pi/4)\mathcal{Y}^{-1}=\nn \mathcal{H}(\pi-\phi_{L},\pi-\phi_{R},\alpha=\pi/2, \beta=\pi/4,3\pi/4).
\end{align} 
This suggests $I(\phi)=-I(\pi-\phi)$ and $I(\phi)=-I(-\phi)$ for Eqs. (\ref{23}) and (\ref{24}), respectively.
This leads us to establish that $\eta$ would be zero for these values of $\beta$ in the junction. 

Now for $\alpha=0$ and $\pi$ Hamiltonian transforms under $\mathcal{Z}$ transformation as,
\begin{align}
    \mathcal{Z}\mathcal{H}(\phi_{L},\phi_{R},\alpha=0,\pi,\beta)\mathcal{Z}^{-1}=\nn \mathcal{H}(-\phi_{L},\pi-\phi_{R},\alpha=0,\pi,\pi-\beta).
    \label{25}
\end{align}
Therefore, the junction Hamiltonian follows relations analogues to Eqs.(\ref{23}) and (\ref{24}) thereby producing the similar current and resulting $\eta=0$ for these values of $\beta$.\\
When we replace $p_x$-wave SC with chiral-$p$ wave SC in the Hamiltonian,  the symmetry transformations mentioned in Eqs. (\ref{17}-\ref{25}) remains same as $\mathcal{Y}\mathcal{H}_{chiral}(\phi_{R})\mathcal{Y}^{-1}=\mathcal{H}(\pi-\phi_{R})$ and $\mathcal{Z}\mathcal{H}_{chiral}(\phi_{R})\mathcal{Z}^{-1}=\mathcal{H}_{chiral}(\pi-\phi_{R})$. As a result, the system with chiral-$p$ wave SC on the right exhibits the same current and efficiency relation as the one with $p_{x}$-wave SC. 

\label{symmetry}

\section{Gate voltage and impurity}
\label{sec5}
\begin{figure}
\centering
\begin{tabular}{c c}
   \includegraphics[width=0.44\linewidth]{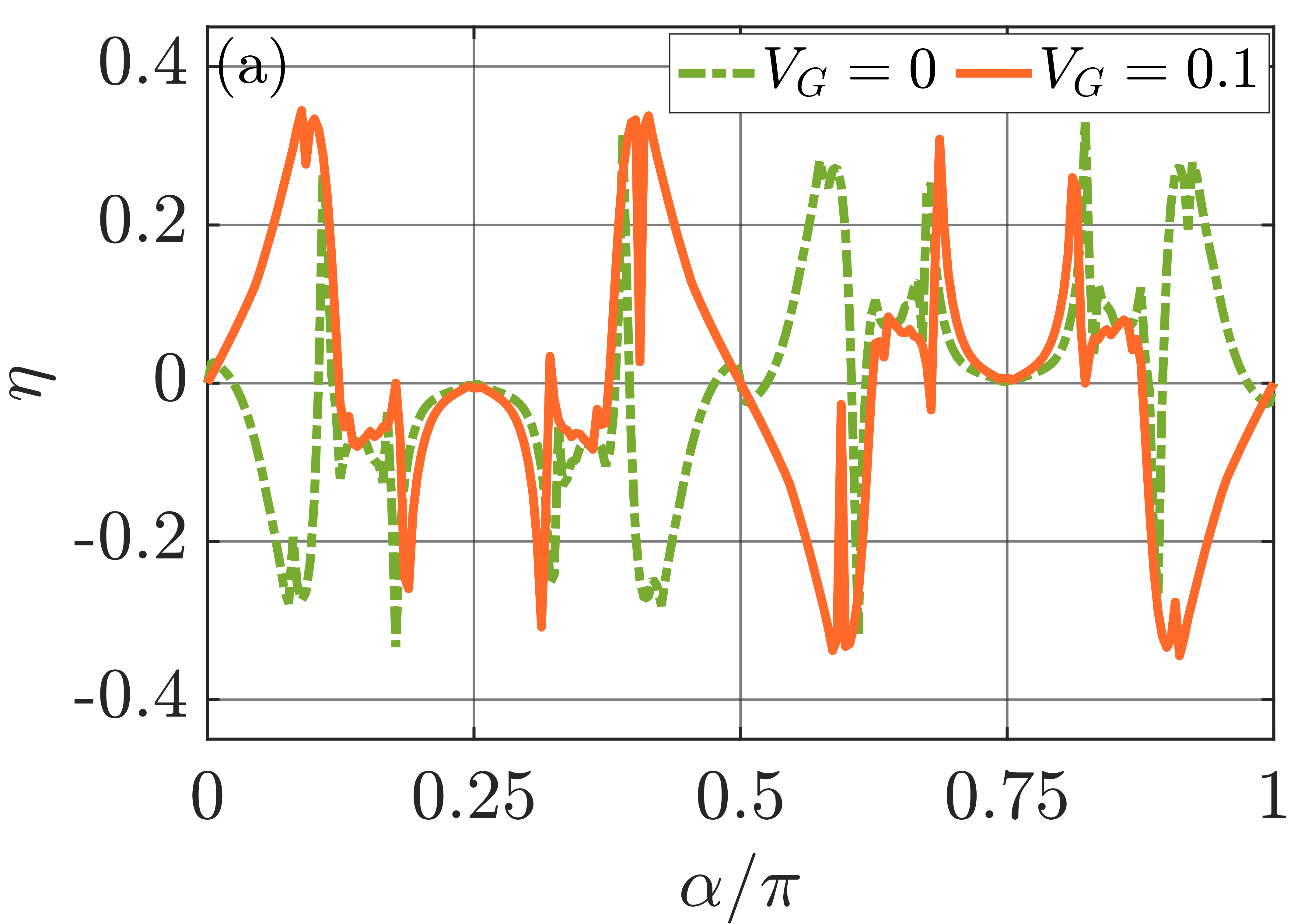}
& \includegraphics[width=0.41\linewidth]{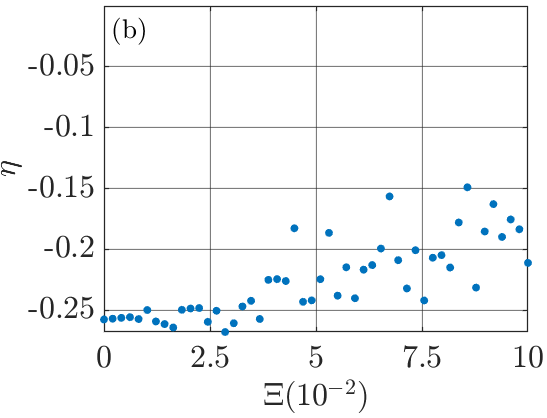}
    \end{tabular}
    \caption{(a) Plot of $\eta$ as a function of $\alpha$ with $V_{G}=0.1$(orange) and $V_{G}=0$(green). (b) Effect of disorder on $\eta$ is shown, disorder is modeled using box distributed random variable $\Xi$ where $\alpha=0.07\pi$. Remaining values of parameters are $N_{x}=4$, $N_{y}=6$, $J_a=0.7$ and $\lambda=0.2$}
\label{fig6} 
\end{figure}

In the previous sections, gate voltage $V_{G}$ in the AM region has been taken zero.  However, gate voltage can modulate the efficiency of system by acting as a potential barrier [\onlinecite{Gupta2023-rw,PhysRevB.109.174511,PhysRevLett.131.096001}].   Thus, we explore the effect of the tunable gate voltage applied in the AM region on the efficiency of the system. We plot the variation of $\eta$ with the lobe angle of AM $\alpha$ for $V_{G}=0.1$. We observe from the Fig. \ref{fig6}(a) that high efficiency can persist for larger range of alpha when $V_{G}=0.1$ than for $V_{G}=0$. Another remarkable feature is that by tuning gate voltage one can switch the diode efficiency from positive to negative. Moreover, not only we can switch the sign of efficiency, but one can increase the efficiency in the junction for same value of $\alpha$. Since the system with the gate voltage obeys the same symmetry transformation as the system without it, we observe that it exhibits the same characteristics discussed in Sec. \ref{sec4}. Next, we examine the effect of disorder on the efficiency of the system and modeled it as the random box disorder potential at onsite in AM region. The range of the box distribution around the onsite potential of clean system, $\Xi$ varies along the $x-$axis in the Fig.\ref{fig6}(b). Here each point is the average over 10 random distribution, averaging over number of multiple disorder realizations helps to reveal the trends while maintaining the effects of randomness. Notable, upto the variance of approx $5\%$ of the clean onsite potential, efficiency doesn't vanish. Thereby, our system shows nonreciprocity and thus JDE even in the presence of disorder.

\section{More systems}
\label{sec6}
To complete our discussion, we explore three more junctions between SCs keeping AM with SOC in between, namely, $p_x$-wave SC/AM/$p_x$-wave SC, $p_x$-wave SC/AM/$p_y$-wave SC, and $d$-wave SC/AM/$d$-wave SC. Among these three junction, $p_x$-wave SC/AM/$p_x$-wave SC shows no diode effect however the other two sustains non-reciprocity but the efficiency is very minute. To understand the non reciprocity, we introduce another mirror symmetry transformation about $yz$ plane, $M_{yz}$. The matrix form of $M_{yz}$ for the BdG Hamiltonian is  given as,
\[
M_{xz}= \begin{pmatrix}
    i\sigma_x & 0\\
    0 & -i\sigma_x
\end{pmatrix}\mathcal{I}_{x},
\]
where $\mathcal{I}_{x}$ is inversion of $x-$axis ($x\rightarrow-x$ and $k_{x}\rightarrow-k_{x}$). Now we introduce a joint symmetry $\mathcal{W}=M_{xz}M_{yz}$ which transforms the Hamiltonian of junction $p_x$-wave SC/AM/$p_x$-wave SC as,
\[
\mathcal{W}\mathcal{H}(\alpha,\phi_{L},\phi_{R})\mathcal{W}^{-1}=\mathcal{H}(\alpha,\pi+\phi_{R},\pi+\phi_{L}).
\]
Since the inversion of $x$-axis also inverts the current with the swapping of SCs on the left and right, thereby $I(\alpha,\phi)=-I(\alpha,-\phi)$. The symmetry transformation $\mathcal{W}$ prohibits this junction to have the non-reciprocity for any value of $\alpha$. Similarly, in $d$-wave SC/AM/$d$-wave SC, current satisfies relation $I(\alpha,\phi,\beta_{L},\beta_{R})= I(\alpha,-\phi,\beta_{R},\beta_{L})$ due to $\mathcal{W}$, where $\beta_{L(R)}$ is lobe angle of the left (right) $d-$wave SC from $x$-axis. This current relation suggests that for non-reciprocity $\beta_{R}\neq \beta_{L}$ should be satisfied.\\

We also numerically compute $\eta$ for junctions where ferromagnet (FM) with SOC is in between the different SCs as above, the system parameters are same as previously mentioned junctions.  The form of Hamiltonian for FM regime with magnetization considered in $z$-direction, has the form, $\hat H_M=\hat H_0(k)+\hat \Lambda(k)+\hat M$ with $\hat H_0(k)$, $\hat \Lambda(k)$ defined in Eqs. (\ref{eq1}), (\ref{eq4}), respectively and $\hat M=m_z \tau_0 \otimes \sigma_z$ with $m_z=0.5t_{0}$. But surprisingly, even in the presence of SOC, broken TRS and inversion CPR comes out to be reciprocal, except when at least one of the SC is $d-$wave. Notably when $s-$ wave SC and ST SC are considered in the junction, non-reciprocity only exists when AM is in between. Conversely, when junction is formed with $d$-wave SC and ST SC, $\eta\neq0$ for both AM and FM. The reason behind this result is that the non-reciprocity requires the anisotropy of spin polarization in the Fermi surface or anisotropy in pair-potential of SC. However, breaking inversion along $x-$axis (either by using non-identical SCs on both side or if same SCs are forming the junctions then different lobe angle of pair potential is required) is necessary in each of such junctions.

\section{Conclusion}
\label{sec7}
We have examined the CPR for a variety of SS/AM/ST JJs that exhibit non-reciprocity even without external magnetic field. The CPR has been calculated using Green's function techniques and M\"obius transformation. We have shown that Rashba SOC is necessary for having JDE but not the sufficient condition, as symmetry transformations satisfied by the system leads to the absence of diode effect. We have shown the existence of high efficiency for the large parameter range of Rashba SOC, AM strength and crystallographic angle of AM. The obtained efficiency can be switched from positive to negative as well as its magnitude can be modulated using a gate potential. Our study in addition to its application in JDE which is stable against disorder, would also help to examine of the role of SOC and mechanisms involved in presence of unconventional magnet leading to non-reciprocity.  We have also done a comparative study of JJs with FM in between, which suggests us four key requirements for such JJs to have non reciprocity in the current.  First requisite is broken TRS, second is having non-identical left and right SC in the JJ (this also includes SC with different angle of lobe direction of pair potential such as in d-wave SC), third is the presence of SOC, and last one is the presence of anisotropy in spin polarization at the Fermi surface or anisotropy in the pair potential of SC.

\textit{Acknowledgments\textemdash}  For financial support, L.S. thanks UGC, India. The authors thank C. Schrade and Bo Lu for stimulating discussions on related topics.


\appendix
\section{Tight Binding Hamiltonian}
\label{AppA}

The tight binding BdG Hamiltonian of system given in Eq.\eqref{Eq6}, can be rewritten as
\begin{align}
    \hat{H}^{\mathcal{Q}}=\sum_{x,y} \Big(\Psi^{\dagger}_{x,y}\hat{H}_{0}^{\mathcal{Q}}\Psi_{x,y} + \Psi_{x+1,y}^{\dagger}\hat{H}^{\mathcal{Q}}_{x}\Psi_{x,y} \nonumber\\+ \Psi_{x,y+1}^{\dagger}\hat{H}^{\mathcal{Q}}_{y}\Psi_{x,y}
+\Psi_{x+1,y+1}^{\dagger}\hat{H}^{\mathcal{Q}}_{xy}\Psi_{x,y} \nonumber\\+
\Psi_{x+1,y-1}^{\dagger}\hat{H}^{\mathcal{Q}}_{x\bar{y}}\Psi_{x,y} + \text{H.c.}\Big),
\end{align}
where $Q\in LS, AM, RS$ represnts left SC, altermagnet, and right SC, respectively. and $\hat{H_x}$ ($\hat{H_{y}}$) is nearest neighbor hopping matrix along $x$ ($y$) direction. The hopping matrices $\hat{H}_{xy}$ and $\hat{H}_{x\bar{y}}$ corresponds to next nearest neighbor, which is present in $d$-wave SC and AM. The spinor is written as,
$\Psi(x,y)=(\hat{\psi}_{x,y,\uparrow} , \hat{\psi}_{x,y,\downarrow}, \hat{\psi}^{\dagger}_{x,y,\uparrow}, \hat{\psi}^{\dagger}_{x,y,\downarrow})^{T}$. Now when LS is $s-$wave SC,
\begin{align}
    \hat{H}^{LS}_{0}=\begin{pmatrix}
        \Big(\frac{4t_0}{a^{2}}-(\mu+V_G)\Big)\sigma_{0} &  \Delta_{0} e^{-i\phi_{L}}i\sigma_{y}\\
        -\Delta_{0} e^{i\phi_{L}}i\sigma_{y} & - \Big(\frac{4t_0}{a^{2}}-(\mu+V_G)\Big)\sigma_{0}
    \end{pmatrix},
\end{align}

\begin{align}
    \hat{H}_{x}^{LS}=\hat{H}_{y}^{LS}=-\frac{t_0}{a^{2}}\sigma_{0}\tau_{z},
\end{align}

When we have $d$-wave SC on the left, 
\begin{align}
        \hat{H}^{LS}_{0}=\Big(\frac{4t_0}{a^{2}}-(\mu+V_G)\Big)\sigma_{0}\tau_{z},
\end{align}

\begin{align}
    \hat{H}^{LS}_{x}= -\hat{H}^{LS}_{y}= &\frac{\Delta_{0}}{a^{2}} \cos{2\beta}\begin{pmatrix}
        0 & -e^{-i\phi_{L}}i\sigma_{y}\\
        e^{-i\phi_{L}}i\sigma_{y} & 0
    \end{pmatrix} \nonumber\\
   & -\frac{t_0}{a^{2}}\sigma_{0}\tau_{z},
\end{align}

\begin{align}
    \hat{H}^{LS}_{xy} = -  \hat{H}^{LS}_{x\bar{y}}= \frac{\Delta_{0}}{2a^{2}}\sin{2\beta}\begin{pmatrix}
        0 & -e^{-i\phi_{L}}i\sigma_{y}\\
        e^{-i\phi_{L}}i\sigma_{y} & 0
    \end{pmatrix},
\end{align}
for central region of AM we have,
\begin{align}
    \hat{H}^{AM}_{0} = \Big(\frac{4t_0}{a^{2}}-\mu\Big)\sigma_{0}\tau_{z},
\end{align}
\begin{align}
    \hat{H}^{AM}_{x} = -\frac{J_a}{a^{2}}\cos{2\alpha}\sigma_{z}\tau_{z} - &\frac{\lambda }{2ia}(\sin{\alpha}\tau_{0}\sigma_{x}+\cos{\alpha}\tau_{z}\sigma_{y}) \nn
    -&\frac{t_0}{a^{2}}\sigma_{0}\tau_{z},
\end{align}

\begin{align}
    \hat{H}^{AM}_{y} = \frac{J_a}{a^{2}}\cos{2\alpha}\sigma_{z}\tau_{z} - &\frac{\lambda }{2ia}(\sin{\alpha}\tau_{z}\sigma_{y} - \cos{\alpha}\tau_{0}\sigma_{x})\nn
    -&\frac{t_0}{a^{2}}\sigma_{0}\tau_{z},
\end{align}

\begin{align}
    \hat{H}^{AM}_{xy} =-\hat{H}^{AM}_{x\bar{y}}=- \frac{J_a}{2a^{2}}\sigma_{z}\tau_{z}\sin{2\alpha},
\end{align}
for right SC,

\begin{align}
    \hat{H}_0^{RS} = \Big(\frac{4t_0}{a^{2}}-\mu\Big)\sigma_{0}\tau_{z},
\end{align}
\begin{align}
    \hat{H}_{x}^{RS} = \begin{pmatrix}
        0&\frac{\eta_{1}\Delta_0}{2ia}e^{-i\phi_{R}}\\
        \frac{\eta_{1}^{*}\Delta_0}{2ia}e^{i\phi_{R}} & 0
    \end{pmatrix} -\frac{t_0}{a^{2}}\sigma_{0}\tau_{z},
\end{align}

\begin{align}
    \hat{H}_{y}^{RS} = \begin{pmatrix}
        0&\frac{\eta_{2}\Delta_0}{2a}e^{-i\phi_{R}}\\
        -\frac{\eta_{2}^{*}\Delta_0}{2a}e^{i\phi_{R}} & 0
    \end{pmatrix} -\frac{t_0}{a^{2}}\sigma_{0}\tau_{z},
\end{align}
Where $a$ is the lattice constant which we have taken unity throughout the paper.

\begin{align}
    H^{C}&=\sum_{y}\Big(\psi^{\dagger}_{x=0,y}\hat{H}_{tL}\psi_{x=1,y}  \nonumber\\
    &+ \psi^{\dagger}_{x=L,y}\hat{H}_{tR}\psi_{x=L+1,y} 
    + \text{H.c.}\Big ),
\end{align}
where,
\begin{align}
    \hat{H}_{t,L(R)}=\begin{pmatrix}
        t_{L(R)}\sigma_0 & 0\\
        0 &-t_{L(R)}^{*}\sigma_0
    \end{pmatrix}.
\end{align}
Here $t_{L(R)}$ is the hopping parameter from left SC to AM and from AM to right SC.

\section{M\"obius Transformation}
\label{mob}
For the surface Green's function we  use M\"obius transformation using M\"obius transformation matrix defined for left and right semi-infinite SC as [\onlinecite{PhysRevB.55.5266}],
\begin{align}
    X_{L}=\begin{pmatrix}
        0 & (H_{12}^{LS})^{-1}\\
        -H_{12}^{LS\dagger} & [(E + i\delta) - H_{11}^{LS}] (H_{12}^{LS})^{-1}
    \end{pmatrix},
\end{align}
and
\begin{align}
    X_{R}=\begin{pmatrix}
        0 & (H_{12}^{RS\dagger})^{-1}\\
        -H_{12}^{RS} & [(E + i\delta) - H_{11}^{RS}] (H_{12}^{RS\dagger})^{-1}
    \end{pmatrix},
\end{align}
Next we consider a matrix $U_{L(R)}$ which diagonalizes $X_{L(R)}$ as $U_{L(R)}^{-1}X_{L(R)}U_{L(R)}=\Lambda_{L(R)}$. The matrix $\Lambda_L/R$ contains all eigenvalues, arranged in ascending order of absolute values. Further,  we write $U_{L(R)}$ in block form as
\begin{align}
    U_{L(R)}=\begin{pmatrix}
        U_{L(R)11} & U_{L(R)12}\\
        U_{L(R)21} & U_{L(R)22}
    \end{pmatrix}.
\end{align}
Finally, the surface Green's function for the left (right) SC is calculated as $g^{r}_{LS(RS)}=U_{L(R)12}U_{L(R)22}^{-1}$.

\bibliography{biblio}

\end{document}